\documentclass[showpacs,aps,pre,twocolumn,amsmath,amssymb,epsfig]{revtex4-1}

\usepackage{graphicx}
\usepackage{epsfig}
\usepackage{dcolumn}
\usepackage{bm}
\newcommand     {\beq}[1]         { \begin{equation} #1 \end{equation} }
\newcommand     {\beqa}[1]        { \begin{eqnarray} #1 \end{eqnarray} }

\begin{document}

\title{Time dependent fracture under unloading in a fiber bundle model}

\author{R\'eka K\"orei}
\author{Ferenc Kun}
\email{Corresponding author: ferenc.kun@science.unideb.hu}
 \affiliation{Department of Theoretical Physics, University of Debrecen,
P.O. Box 5, H-4010 Debrecen, Hungary}

\begin{abstract}
We investigate the fracture of heterogeneous materials occurring under 
unloading from an initial load. Based on a fiber bundle model of time dependent 
fracture, we show that depending on the unloading rate the system has two phases:
for rapid unloading the system suffers only partial failure and it has an infinite 
lifetime, while at slow unloading macroscopic failure occurs in a finite time. 
The transition between the two phases proved to be analogous to continuous 
phase transitions. Computer simulations revealed that during unloading 
the fracture proceeds in bursts of local breakings triggered by 
slowly accumulating damage. 
In both phases the time evolution starts with a relaxation of the bursting activity 
characterized by a universal power law decay of the burst rate. 
In the phase of finite lifetime the initial slowdown is followed by an acceleration 
towards macroscopic failure where the increasing rate of bursts obeys the 
(inverse) Omori law of earthquakes. We pointed out a strong correlation 
between the time where the event rate reaches a minimum value and of the lifetime 
of the system which allows for forecasting of the imminent 
catastrophic failure. 
 \end{abstract}

\maketitle

\section{Introduction}
In applications, materials are often subject to constant or slowly varying 
external loads which fall below their fracture strength. 
Such sub-critical loads usually give rise to time dependent deformation 
and failure in a finite time 
\cite{herrmann_statistical_1990,busse_damage_2006,alava_statistical_2006}. 
Creep rupture (constant load)
and fatigue failure (varying load) set serious limitations on the applicability 
of materials in construction components, and they are often responsible 
e.g.\ for the instability of steep slopes in mountains leading to the emergence of 
natural catastrophes such as landslides and collapse of rock walls 
\cite{landslides_book_deblasio}. 
Sub-critical fracture of heterogeneous materials 
proceeds in bursts of local rupture events, which can be registered in the form
of acoustic signals. Acoustic emission measurements provide a valuable insight
into the dynamics of fracture processes addressing also the possibility 
of forecasting the imminent final collapse 
\cite{guarino_experimental_1998,nataf_avalanches_2014}.

Failure of materials can also occur due to unloading from a previously applied 
stress level. Excavation during underground engineering rapidly releases stress
which can result in rock bursts \cite{wang_experimental_2015}.
 Similar conditions may also occur on 
much larger length and time scales at the emergence of earthquakes: crustal unloading 
due to near-surface mass redistribution (water, ice or quarried material) 
can affect the subsurface stress field, altering seismic activity and being 
also responsible for rupture activation and induced earthquakes 
\cite{gonzalez_2011_2012}. Fracture processes under unloading 
present a high degree of complexity, which makes it difficult to achieve a general
understanding.

To consider this problem, in the present paper we investigate the process 
of sub-critical fracture, which occurs when unloading from an initial 
load. We use a fiber bundle model of time dependent deformation and rupture, 
which captures the slow damaging of loaded fibers and their immediate breaking 
when the local load exceeds the fibers' fracture strength. 
In the model, we focus on the case when a constant sub-critical load gives rise 
to failure in a finite time so that unloading may prevent the final breakdown.
We show by analytical calculations and computer simulations that 
the system has two phases, i.e.\ at rapid unloading only partial failure occurs
and the sample has an infinite lifetime. However, slow unloading results in 
global failure in a finite time. We demonstrate that the transition between 
the phases of finite and infinite lifetime occurs at a well-defined unloading 
rate, and it is analogous to continuous phase transitions. 
The unloading process is accompanied by breaking bursts of fibers with 
a varying rate. We show by computer simulations that in the regime of 
finite lifetime the initial relaxation is followed by a short acceleration 
period towards failure. Based on the pattern of the time varying burst rate
we propose a method to forecast the impending failure under unloading.

\section{Fiber bundle model of time dependent fracture}
To investigate the effect of unloading on the process of sub-critical fracture 
we use a fiber bundle model (FBM), which has been introduced recently for the time dependent 
fracture of heterogeneous materials 
\cite{1742-5468-2009-01-P01021,kun_universality_2008,PhysRevE.85.016116}.
In the model, the sample is represented by a parallel set of fibers which can 
only be loaded
along the fibers' direction. The fibers have linearly elastic behavior with a
constant Young modulus $E$. Subjecting the bundle to a constant load $\sigma_0$ 
below the fracture strength $\sigma_c$ of the system, the fibers break due to two 
physical mechanisms: immediate breaking occurs when the local load $\sigma_i$ 
on fibers exceeds their fracture strength 
$\sigma_{th}^{i}$, $i=1,\ldots , N$, where $N$ is the number of fibers. 
Under a sub-critical load $\sigma_0<\sigma_c$ this breaking mechanism would 
lead to a partially failed configuration with an infinite lifetime.
Time dependence arises such that those fibers, which remained intact
under a given load, undergo an aging process accumulating damage $c(t)$. 
We assume that the rate $\Delta c_i$ of damaging has a power law dependence 
on the local load $\sigma_i$ ($i=1,\ldots , N$)
\begin{eqnarray}
\Delta c_i = a\sigma_i^{\gamma}\Delta t,
\label{eq:damlaw}
\end{eqnarray}
where $a$ is a constant and the exponent $\gamma$ controls the characteristic 
time scale of the aging process with $1\leq \gamma < +\infty$. 
The total amount of damage $c_i(t)$ accumulated up to time $t$ can be obtained
by integrating over the entire loading history of fibers 
$c_i(t)=a\int_0^t\sigma_i(t')^{\gamma}dt'$. 
Fibers can sustain only a finite amount of 
damage so that when $c_i(t)$ exceeds the local damage threshold $c^i_{th}$, 
the fiber breaks.

After each breaking event, the load of the failed fiber gets redistributed over 
the remaining intact ones. In the present study we focus on the case of equal load sharing 
(ELS), i.e.\ when a fiber breaks its load is equally shared by all remaining fibers. 
ELS has the consequence that no stress fluctuations can emerge 
in the system, since all the fibers keep the same load. 
Heterogeneity of the material is entirely represented by the randomness of the 
failure thresholds of fibers $\sigma_{th}^i, c_{th}^i$, $i=1,\ldots , N$ of 
the two breaking modes. For simplicity, both thresholds were sampled 
from a uniform distribution between zero and one $0<\sigma_{th}^i \leq 1$, 
and $0<c_{th}^i \leq 1$ without any correlation.

It has been demonstrated that even in the simplest case of a constant sub-critical 
load $\sigma_0$ a highly complex
fracture process emerges: when the load is set weak fibers $\sigma_{th}^i<\sigma_0$ 
break immediately \cite{kun_fatigue_2007,kun_universality_2008}. The load of broken fibers gets 
redistributed over the intact ones, which may induce further breaking events followed again
by load redistribution. As the consequence of successive breaking and redistribution steps, 
an avalanche is triggered which stops when all the remaining intact fibers are 
strong enough to keep the external load. The time evolution of the bundle starts from this
partially fractured initial state. Loaded fibers accumulate damage and break slowly, 
one-by-one due to damaging. Since damage breakings are also followed by load redistribution,
they gradually increase the load on intact fibers, and in turn, can trigger sudden bursts 
of immediate breaking \cite{kun_fatigue_2007,kun_universality_2008}. 
Eventually, the time evolution of the fracture process
sets in as a series of sudden bursts, analogous to acoustic outbreaks in real experiments,
separated by quite periods of slow damaging.
In the model, loaded fibers always accumulate damage, and eventually 
break, hence, at any finite load $\sigma_0  > 0$ which is kept constant, 
the system has a finite lifetime $t_f$.
It has been shown in Refs.\ \cite{kun_fatigue_2007,1742-5468-2009-01-P01021} 
that in the sub-critical regime $\sigma_0<\sigma_c$ 
the lifetime $t_f$ of the bundle decreases as a power law of $\sigma_0$, while super-critical
loads $\sigma_0 > \sigma_c$ give rise to immediate global failure.

The model has been successfully applied
to describe the time evolution of damage induced creep rupture under 
a constant load \cite{kun_fatigue_2007,1742-5468-2009-01-P01021,danku_creep_2013}, 
the statistics of crackling bursts 
\cite{kun_universality_2008,PhysRevE.85.016116,danku_frontiers_2014}, 
the average temporal profile  \cite{danku_PhysRevLett.111.084302},
and the fractal geometry of bursts \cite{PhysRevE.92.062402} emerging 
during the fracture process.

\section{Time evolution of damage}
\label{sec:damage}
In order to investigate the fracture process of the bundle under unloading,
we assume that the externally applied load decreases with time according to 
a linear profile
\beq{
\sigma(t) = \sigma_0 - At,
}
where $\sigma_0 =\sigma(t=0)$ denotes the initial sub-critical load and the parameter $A>0$ 
is the rate of unloading. 
For simplicity, we start the analysis with the case when damage accumulation is the only breaking 
mechanism of fibers so that no avalanches of immediate breaking can be triggered. 
It has the advantage that the most important characteristics of the time evolution 
of the bundle can be deduced analytically.

During the time evolution of the system the degree of degradation of the bundle 
can be quantified by the fraction of broken fibers $n_b(t)=N_b(t)/N$, 
where $N_b(t)$ denotes the total number of fibers broken
up to time $t$. The rate of breaking $dn_b/dt$ can be obtained from the damage 
law Eq.\ (\ref{eq:damlaw}) as 
\begin{eqnarray}
\frac{dn_b(t)}{dt} = af(c_{th})\sigma_s(t)^\gamma, \label{eq:dnbdt1}
\end{eqnarray}
where $f(c_{th})$ denotes the probability density function of damage thresholds.
Since equal load sharing is assumed, the load $\sigma_s(t)$ of single fibers at time $t$ 
reads as
\beq{
\sigma_s(t) = \frac{\sigma_0 - At}{1 - n_b(t)}, \label{eq:pt1}
}
where the nominator and the denominator take into account the decreasing 
externally applied stress, and the reduction of the load bearing cross section
of the bundle, respectively. The above expressions demonstrate that the 
time evolution of the bundle is controlled by two competing mechanisms, 
namely, the decreasing external load favors the slowdown of fracturing, 
however, fiber breaking gives rise to the increase of the load on single 
fibers, which accelerates the failure process. The competition is controlled
by the rate of unloading $A$. 

For uniformly distributed damage thresholds, with the density function 
$f(c_{th})=1$, the evolution equation Eq.\ (\ref{eq:dnbdt1})
can be solved analytically 
\begin{equation}
n_b = 1-\left[1-\frac{a \sigma_0^{\gamma+1}}{A} \left \lbrace 1 
- \left( 1-\frac{A}{\sigma_0}t \right)^{\gamma+1}\right \rbrace 
\right]^{\frac{1}{\gamma +1}}, \label{eq:nb}
\end{equation}
which fulfills the initial condition $n_b(t=0)=0$.  
\begin{figure}
\begin{center}
\epsfig{bbllx=40,bblly=40,bburx=375,bbury=330,file=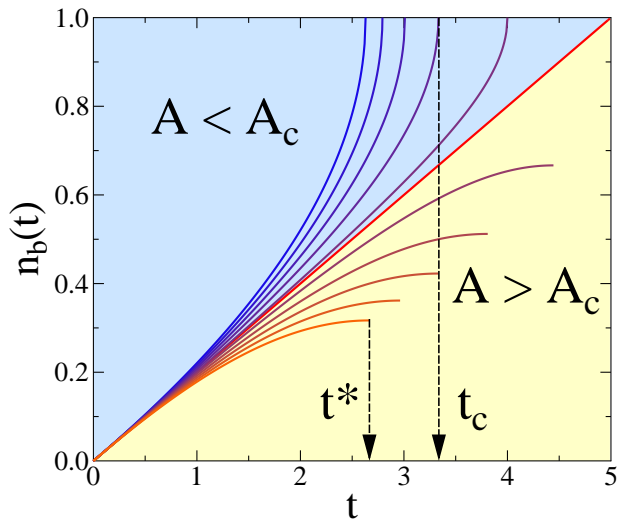, width=8.3cm}
\end{center}
  \caption{(Color online) The fraction of broken fibers $n_b$ as a function of time $t$
  for several values of the unloading rate $A$ at a fixed initial load 
  $\sigma_0/\sigma_c=0.01$ and $\gamma=2$.
  Two qualitatively different regimes can be identified: for slow unloading global 
  failure emerges in a finite time $t_c(A)$, while for rapid unloading only a partial 
  failure occurs and the system has an infinite lifetime. $t^*(A)$ and $t_c(A)$,
  given by Eqs.\ (\ref{eq:tstar},\ref{eq:t_c}),
  indicate the end of the time evolution where either the external 
  load drops down to zero or all the fibers break, respectively.
  \label{fig:nb}
}
\end{figure}
The $n_b(t)$ curves 
are presented in Fig.\ \ref{fig:nb} for the damage accumulation exponent $\gamma=2$ 
at several values of the unloading rate $A$,
keeping the initial load fixed $\sigma_0/\sigma_c=0.01$. Depending on the value of $A$, 
two qualitatively different regimes of the time evolution can be identified 
in the figure: 
for slow unloading the fraction of broken fibers grows with an increasing rate and 
reaches one at a finite time $t_c$, where global failure of the bundle occurs. 
The reason is that for low $A$ the fibers are loaded for a sufficiently long time 
to accumulate enough damage to break. This mechanism implies that for 
decreasing $A$ the lifetime $t_c$ of the bundle decreases, and in the limit 
of $A\to 0$ it tends to the lifetime of constant external load 
$t_c=a\sigma_0^{-\gamma}$ obtained in Refs.\ \cite{kun_fatigue_2007,1742-5468-2009-01-P01021}. 
For rapid unloading the fibers can accumulate only a lower amount of damage, 
and hence, by the time 
\beq{
t^*=\sigma_0/A,
\label{eq:tstar}
}
when the external load reaches zero, a finite fraction of fibers survives. Consequently,
the $n_b$ curves approach their limit values $n_b(t^*)<1$ with a decreasing rate.
It has to be emphasized that in this regime the time evolution of the system 
terminates at $t^*$, but the partially failed state attained eventually 
has an infinite lifetime.

\begin{figure}
\begin{center}
\epsfig{bbllx=30,bblly=5,bburx=370,bbury=300,file=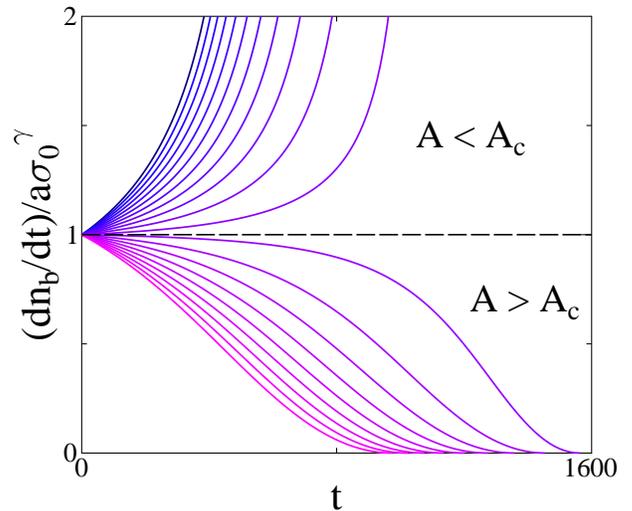, width=8.3cm}
\end{center}
  \caption{(Color online) Evolution of the breaking rate of fibers at different
  values of the unloading rate $A$ for the same parameters as in Fig.\ \ref{fig:nb}.
  At the critical unloading rate $A_c$ the breaking rate is constant 
  $dn_b(t)/dt=a\sigma_0^{\gamma}$, which was used for normalization.
  \label{fig:dnbdt}
}
\end{figure}

Substituting Eq.\ (\ref{eq:nb}) into Eq.\ (\ref{eq:dnbdt1}) the time dependent 
breaking rate can be cast into the final form
\beq{
\frac{dn_b(t)}{dt} = \frac{a(\sigma_0 - At)^{\gamma}}{\left[1-
\frac{a \sigma_0^{\gamma+1}}{A} \left \lbrace 1 
- \left( 1-\frac{A}{\sigma_0}t \right)^{\gamma+1}\right \rbrace 
\right]^{\frac{\gamma}{\gamma +1}}}, \label{eq:dnbdt}
}
which is illustrated in Fig.\ \ref{fig:dnbdt} for several values of $A$. 
It can be observed that the behavior 
of the breaking rate confirms the existence of two phases, i.e.\ for slow 
unloading the system accelerates towards global failure and has a finite lifetime $t_c$,
while for fast unloading partial failure is approached by slowdown and the system has 
an infinite lifetime. The transition between the two phases occurs at a well-defined 
critical unloading rate $A_c$.

The lifetime of the bundle $t_c$ can be obtained as the time where the breaking rate 
Eq.\ (\ref{eq:dnbdt}) diverges
\begin{equation}
t_c = \frac{\sigma_0}{A} \left[ 1- \left(1-\frac{A}{a\sigma_0^{\gamma+1}} 
\right)^{\frac{1}{\gamma+1}} \right].  \label{eq:t_c}
\end{equation}
It can be seen that $t_c$ is only finite if the unloading rate $A$ falls below
the characteristic value
\beq{
A_c=a\sigma_0^{\gamma + 1}, \label{eq:a_c}
}
which defines the critical point $A_c$, separating the finite and infinite 
lifetime regimes of the 
system. The critical unloading rate depends on the initial load $\sigma_0$ 
and on the exponent 
$\gamma$, which controls the load dependence of the aging process.

\begin{figure}
\begin{center}
\epsfig{bbllx=40,bblly=40,bburx=375,bbury=330,file=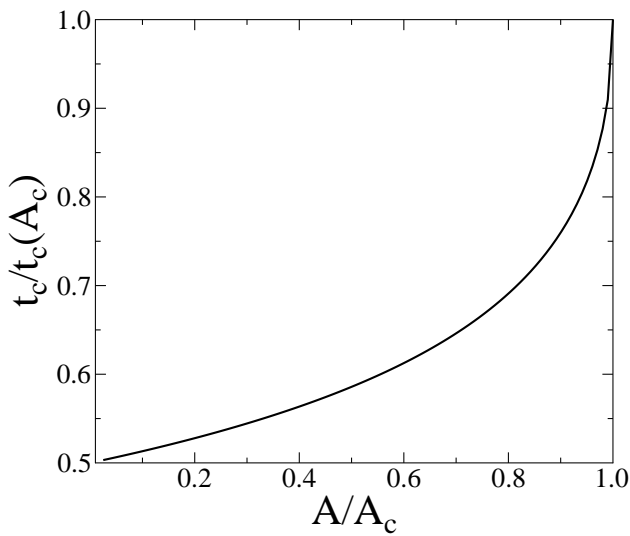, width=8.3cm}
\end{center}
  \caption{The lifetime of the bundle as a function of the control parameter $A$
  below the critical point $A\leq A_c$, normalized by the maximum lifetime reached 
  at the critical unloading rate $t_c(A_c)$. 
  \label{fig:tc}
}
\end{figure}

It is a very interesting question how the transition between the two phases occurs, 
when the control parameter $A$ is varied.
Starting from Eq.\ (\ref{eq:t_c}), it can be simply shown that in the limiting case
of $A\to 0$ the value of $t_c$ converges to $t_c(A=0)=\sigma_0^{-\gamma}/a(\gamma + 1)$, 
which is the 
lifetime of the bundle under the constant load $\sigma_0$ \cite{1742-5468-2009-01-P01021}. 
For increasing unloading 
rate $A$, as the critical point $A_c$ is approached from below, $t_c$ increases 
and converges to a finite maximum $t_c=\sigma_0/A_c = \sigma_0^{-\gamma}/a$. 
The behavior of $t_c$ as a function of the unloading rate Eq.\ (\ref{eq:t_c}) 
is illustrated in Fig.\ \ref{fig:tc}. 
It follows from Eq.\ (\ref{eq:t_c}) that although $t_c$ remains finite at $A_c$, 
its derivative exhibits a power law divergence as a function of the 
distance $A_c-A$ from the critical point
\begin{equation}
\frac{dt_c}{dA} \sim \left( A_c-A \right)^{-\alpha}.
\end{equation}
The critical exponent $\alpha$ depends only on the damage accumulation exponent
$\alpha=\gamma/(\gamma + 1)$.
Above the critical point $A>A_c$,
the lifetime becomes infinite $t_c\to\infty$ because the unloading 
is so fast that some fraction of fibers remains intact by the time $t^*=\sigma_0/A$,
when the external load reaches zero. 

\begin{figure}
\begin{center}
\epsfig{bbllx=15,bblly=15,bburx=350,bbury=300,file=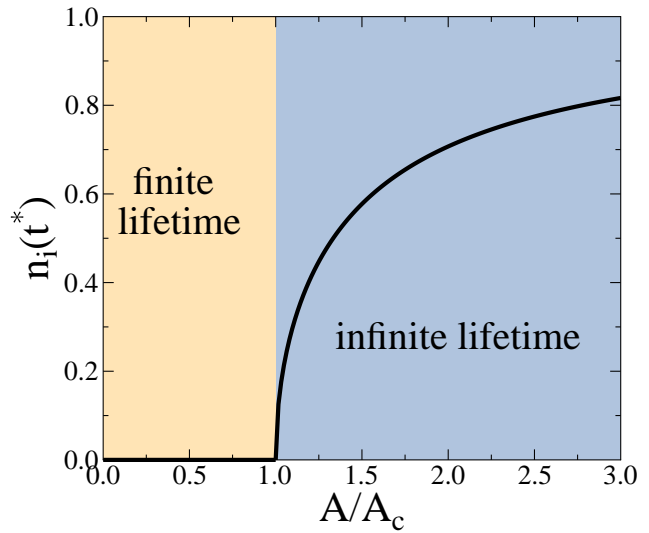, width=8.3cm}
\end{center}
  \caption{{\it (Color online)} The fraction of intact fibers $n_i(A,t^*)$ at the end 
  of the unloading process for the same parameters as before.
  This quantity plays the role of the order parameter of the transition, since 
  it is identically zero in the phase of finite lifetime, while it takes finite non-zero
  values in the phase of partial failure. Approaching the critical point from above 
  $n_i(A,t^*)$ goes to zero as a power law of $A-A_c$ according 
  to Eq.\ (\ref{eq:ni_crit}).
  \label{fig:ni}
}
\end{figure}

The final state of the time evolution 
can be characterized by the fraction of intact fibers $n_i(A,t=t^*)=1-n_b(A, t=t^*)$ 
at the end of the unloading process. 
In the regime of finite lifetime $t_c(A)<t^*(A)$ holds, which implies 
that $n_i(A,t^*)=0$ for $A\leq A_c$ since the failure of the bundle has been 
completed. Above the critical point $A>A_c$ partial failure occurs 
giving rise to a finite
fraction of surviving fibers $n_i(A,t^*)>0$. It follows from Eq.\ (\ref{eq:nb}) 
that $n_i(A,t^*)$ increases with the distance from $A_c$ according to a power law 
\beq{
n_i(A,t^*)\sim (A-A_c)^{\beta},
\label{eq:ni_crit}
}
where the value of the critical exponent is $\beta=1/(\gamma + 1)$. 
The above quantitative analysis shows that the transition between 
the phases of finite and infinite lifetime occurs at $A_c$ analogous 
to a continuous phase transition \cite{nishimori_phase_transition}. 
The fraction of intact fibers $n_i(A,t^*)$
can be considered as the order parameter of the transition, which is
illustrated in Fig.\ \ref{fig:ni}. 

\section{Breaking bursts triggered by damage sequences}
When both breaking mechanisms, i.e.\ immediate breaking and slow damage accumulation, 
are turned on, the fracture process becomes even more complex: as fibers break due 
to damage the load on intact fibers gradually increases. Such load increments 
may be sufficient to induce immediate breaking of fibers, which in turn can generate 
a cascade of immediate breaking. Due to the interplay of slow damaging and immediate 
breaking of fibers, the fracture process is composed of sudden bursts (avalanches) of fiber 
breakings, which are triggered by slowly evolving damage sequences. The complexity 
of this fracture process can only be explored by means of computer simulations.

\subsection{Computer implementation}
Global load sharing implies that both the load of single fibers $\sigma_s(t)$
and the damage $c(t)$ accumulated up to time $t$ are the same for all the fibers.
It has the consequence that for both breaking mechanisms fibers break in the increasing
order of their respective breaking thresholds. Since the system has only 
quenched disorder, the computer implementation 
of the model starts with sorting the randomly generated breaking thresholds 
$\sigma^i_{th}$ and $c^i_{th}$ ($i=1,\ldots , N$) into increasing order.

During the fracture process, after each damage induced breaking 
it has to be checked whether the resulting load increment 
is sufficient to trigger immediate failure of fibers.
If the bundle remained stable then the next fiber will break again due to damage accumulation.
The time $t_{i+1}$ of the next damage breaking with the damage threshold $c^{i+1}_{th}$
can be obtained analytically from the integral of the damage law Eq.\ (\ref{eq:damlaw})
\beq{
c^{i+1}_{th}-c^i_{th} = a\int_{t_i}^{t_{i+1}}\left[ \frac{N(\sigma_0-At)}{N-N_b(t_i)}
\right]^{\gamma} dt,
}
where $c^i_{th}$ is the threshold of the previous damage breaking which occurred 
at time $t_i$. Here $N_b$ denotes the total number of broken fibers at time $t_i$,
which remains constant from the time $t_i$ to $t_{i+1}$. 
After the damaging fiber has been removed from the bundle, its load gets redistributed 
increasing the load on intact fibers to 
\beq{
\sigma_s(t_{i+1})=\frac{N(\sigma_0-At_{i+1})}{N-N_b(t_{i+1})},
}
where $N_b(t_{i+1})= N_b(t_i)+1$. If the load increment induces immediate breakings,
an avalanche may emerge through  breaking and load redistribution steps, which stops
when all remaining fibers are strong enough to keep the enhanced load. 
The size of an avalanche $\Delta$ is defined as the total number of fibers 
breaking in the failure cascade triggered by the preceding damage event.
It is important
that the avalanche propagation is considered to be much faster than damage accumulation
that's why no time is associated to the avalanche duration.  
The separation of time scales of damaging and bursting has the consequence that 
the fracture of the bundle proceeds as a series 
of bursts of immediate breakings triggered by damage sequences. In the simulations
the temporal evolution of the system is followed until either all fibers break $N_b=N$, or 
the external load decreases to zero $\sigma_0-At^*=0$. 
This technique made us possible to perform computer simulations of the system
varying the number of fibers up to $N=10^6$ fibers.

\subsection{Finite size scaling}
To explore the complexity of the breaking process, simulations were carried out
varying the parameters of the model in broad ranges.
The analytic results of the damage limit of the model in Section \ref{sec:damage} were 
obtained for an infinitely large system. 
Simulations at finite system sizes revealed that breaking avalanches do 
not alter the global qualitative behavior of the system, only quantitative 
details of the fracture process change. 
\begin{figure}
\begin{center}
\epsfig{bbllx=20,bblly=20,bburx=780,bbury=700,file=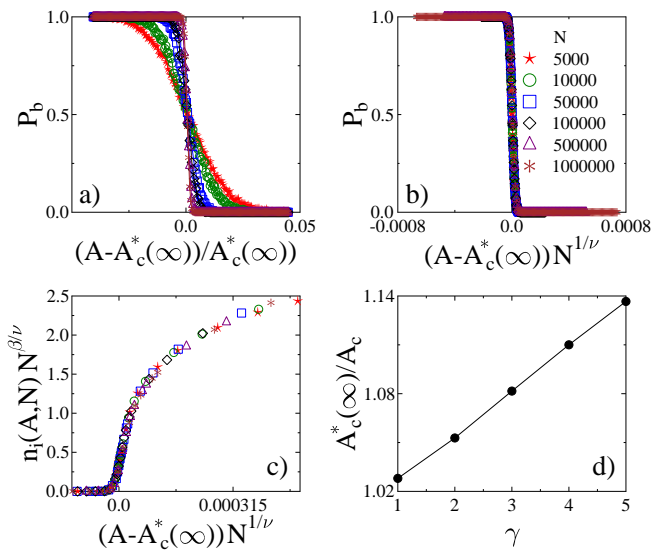, width=8.3cm}
\end{center}
  \caption{{\it (Color online)} $(a)$ Failure probability $P_f$ as a function 
  of $A$ for different system sizes $N$ with $\gamma=2$.
  $(b)$ Rescaling the data of $(a)$ according to Eq.\ (\ref{eq:pf}) a high quality 
  data collapse is obtained. $(c)$ Data collapse of the order parameter 
  obtained at different system sizes using the scaling ansatz 
  Eq.\ (\ref{eq:niscale}). $(d)$ The ratio of the critical values of the unloading 
  rate $A_c^*(\infty)$
  and $A_c$ obtained in the presence and absence of avalanches as a function of the 
  damage accumulation exponent $\gamma$. 
  \label{fig:pf}
}
\end{figure}

Since in avalanches a large number of fibers can break simultaneously, 
avalanche triggering gives rise to an acceleration of the fracture process: 
at the same value of the control parameter $A$ the lifetime $t_c$ of the bundle becomes 
shorter, while the critical point $A_c$ shifts to a higher value compared 
to the analytic results of Eqs.\ (\ref{eq:t_c},\ref{eq:a_c}). 
Higher critical point means that even for faster unloading global failure can 
occur due to the presence of avalanches. In the following the critical point of the system
in the presence of avalanches will be denoted by $A_c^*$ to distinguish it 
from the corresponding value of the purely damage driven case $A_c$.

In order to understand how the value of $A_c^*$ 
depends on the system size, we determined the probability 
of global failure $P_f$ as a function of $A$ varying the number of fibers
$N$ in a broad range. 
For each system size $N$ the value of $P_f$ was obtained
as the fraction of samples which suffered global failure repeating the simulations 
$10^3$ times with different realizations of disorder. 
Figure \ref{fig:pf}$(a)$ shows that for small $N$ failure can occur with a considerable
probability in a broad range of the control parameter $A$.
As the system size increases the transition becomes sharper and a well-defined critical
point emerges. Figure \ref{fig:pf}$(b)$ presents the same data rescaled 
with the system size  using the finite size scaling form of continuous 
phase transition
\beq{
P_f(A,N) = \Psi((A-A_c^*(\infty))N^{1/\nu}).
\label{eq:pf}
}
Here $A_c^*(\infty)$ denotes the critical point of the infinite system $N\to \infty$ 
in the presence of avalanche triggering, while the exponent $\nu$ is the correlation 
length exponent of the transition. In the figure best collapse is achieved with $\nu=2$.

The order parameter of the transition $n_i(A,N)$, i.e.\ the fraction of intact 
fibers at the end of unloading, has a similar dependence 
on the system size $N$. Figure \ref{fig:pf}$(c)$ demonstrates that rescaling 
the data of $n_i(A,N)$ with the number of fibers $N$ according to the scaling form
\beq{
n_i(A,N) = N^{-\beta/\nu}\Phi((A-A_c^*(\infty))N^{1/\nu})
\label{eq:niscale}
}
curves of different system sizes can be collapsed on a master curve. Good quality 
data collapse is obtained using the same value $\nu=2$ of the correlation length exponent
as for the failure probability $P_f(A,N)$. 
For the ratio of the order parameter and correlation length exponents the 
value $\beta/\nu=0.15(6)$ was obtained, which gives $\beta \approx 0.31$ consistent 
with the analytic value $\beta=1/3$ of the damage driven case for $\gamma=2$ used in
Fig.\ \ref{fig:pf}$(c)$. 

Based on the above finite size scaling analysis we determined the asymptotic value of the 
critical unloading rate $A_c^*(\infty)$ for several values of the $\gamma$ exponent 
of the damage accumulation law. In Figure \ref{fig:pf}$(d)$ the ratio of $A_c^*(\infty)$ 
and the analytic value of $A_c$ is presented as a function $\gamma$. 
It can be observed that the critical point
is always higher when avalanches are triggered and the difference from the 
damage driven case increases with $\gamma$.
\begin{figure}
\begin{center}
\epsfig{bbllx=10,bblly=10,bburx=780,bbury=700,file=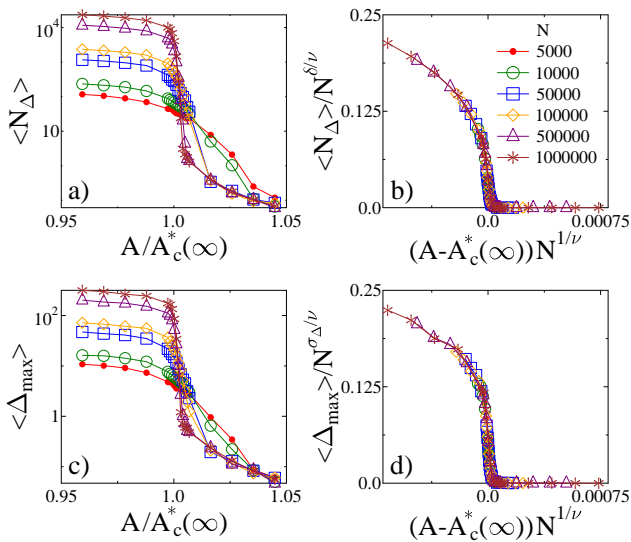, width=8.3cm}
\end{center}
  \caption{{\it (Color online)} The average number of avalanches $(a)$ and the average
  size of the largest avalanche $(c)$ as function of the control parameter $A$
  for different system sizes $N$. Rescaling the curves in $(b)$ and $(d)$ good quality
  data collapse is achieved. The legend of the figures is given in $(b)$.
  \label{fig:max_num_aval}
}
\end{figure}

To quantify the role of breaking avalanches in the fracture process,
we calculated the average total number of avalanches $\left<N_{\Delta}\right>$ 
and the average size 
of the largest avalanche $\left<\Delta_{max}\right>$ as a function of $A$ 
covering both phases.
Figures \ref{fig:max_num_aval}$(a)$ and $(c)$ show that both quantities 
are decreasing functions of the unloading rate $A$, i.e.\ avalanches of largest 
number and size
are triggered when the external load $\sigma$ is kept constant $A=0$, however, unloading 
at any rate reduces the avalanche activity. Similar to the behavior of $P_f(A,N)$ 
and $n_i(A,N)$, these quantities also show the sharpening of the transition between
the phases of finite and infinite lifetime with increasing $N$.
Rescaling the data with the number of fibers 
according to the finite size scaling form Eq.\ (\ref{eq:niscale}),
the curves of different system sizes can be collapsed on the top of each other. 
The exponents $\sigma_{\Delta}$ and $\delta$, controlling the cutoff size and number 
of avalanches as the critical point is approached from below, are defined as
\beqa{
\left<\Delta_{max}\right> &\sim& (A_c^*-A)^{\sigma_{\Delta}},\\
\left<N_{\Delta}\right> &\sim& (A_c^*-A)^{\delta}.
}
Using the same value of $\nu$ as before, and tuning the new exponents
to achieve best collapse, we obtained the values $\delta=1.6(8)$ 
and $\sigma_{\Delta}=1.0(4)$.
Simulations revealed that the correlation length exponent $\nu$, 
and the exponents describing the avalanche activity $\delta$ and $\sigma_{\Delta}$,
are universal, they do not depend on the parameters of the model. 
These exponents are controlled by the range 
of load redistribution, which is fixed here by the equal load sharing rule 
\cite{hansen2015fiber,hidalgo_avalanche_2009,kun_universality_2008}. 
This is also the reason why the size distribution of bursts $p(\Delta)$ exhibits 
the usual power law behavior $p(\Delta) \sim \Delta^{-\tau}$ with the universal 
exponent $\tau=5/2$ (not shown in figures). Since the exponents $\sigma_{\Delta}$ 
and $\delta$ control the cutoff and the integral of the burst size distribution, respectively,
they satisfy the scaling relation $\tau\sigma_{\Delta}=1+\delta$ \cite{danku_disorder_2016}. 
Substituting the above numerical values, our exponents approximately fullfill the relation. 

\section{Rate of breaking bursts}
\begin{figure}
\begin{center}
\epsfig{bbllx=0,bblly=0,bburx=350,bbury=310,file=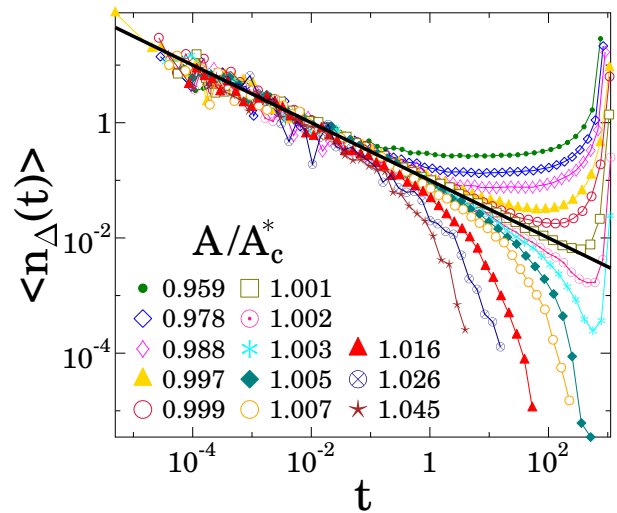, width=8.3cm}
\end{center}
  \caption{{\it (Color online)} Rate of bursts in a system of size $N=10^6$ fibers
  with $\gamma=2$ for several values of the unloading rate $A$. The straight line 
  represents a power law of exponent $-1/2$.
  \label{fig:burstrate}
}
\end{figure}
In order to characterize how the bursting activity evolves at a given value of 
the unloading rate $A$,
we determined the average rate of bursts $\left<n_{\Delta}\right>$ as a function 
of time. Figure \ref{fig:burstrate} presents the event rate for a system of $10^6$
fibers at several values of $A$ below and above the critical point $A_c^*$.
It is interesting to note that in both phases the unloading process is initially 
accompanied by a decreasing rate of bursts. A power law functional form is evidenced 
\beq{
\left<n_{\Delta}\right> \sim t^{-\kappa},
}
where the value of the exponent $\kappa=1/2$ proved to be independent of the initial load $\sigma_0$,
of the unloading rate $A$, and of the damage accumulation exponent $\gamma$. 
In the phase of partial failure the power law decay of the burst rate is followed 
by a rapid exponential decrease, while in the phase finite lifetime 
$\left<n_{\Delta}\right>$ reaches 
a minimum at a time $t_{m}$ which is then followed by an acceleration towards failure.

The curves of Figure \ref{fig:burstrate} were obtained by logarithmically binning 
the time $t$ starting from the beginning of the process, which implies finer 
bins at early times and coarser
ones close to failure. To improve the resolution of the analysis in the close 
vicinity of failure, we re-binned the data of the phase of finite lifetime 
in terms of the time measured from the critical 
point $t_c-t$, using again logarithmic binning. This way we can zoom on 
the acceleration regime of the time evolution, however, the minimum 
of the event rate and the entire relaxation phase, dominating 
in Fig.\ \ref{fig:burstrate}, fall in the last few bins.
\begin{figure}
\begin{center}
\epsfig{bbllx=0,bblly=0,bburx=350,bbury=310,file=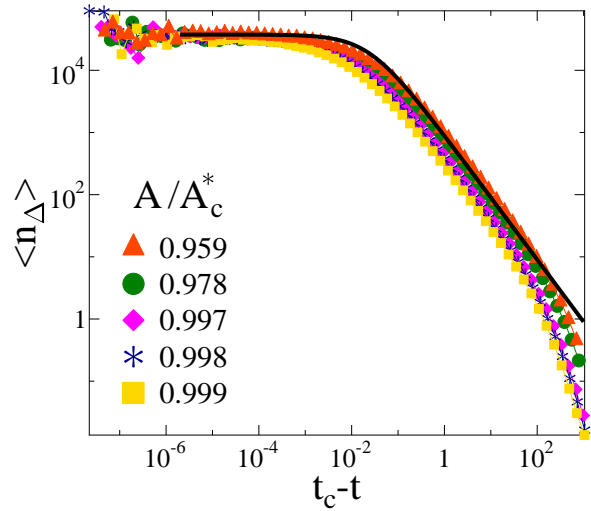, width=8.3cm}
\end{center}
  \caption{{\it (Color online)} Rate of bursts in the same system as in Fig.\ \ref{fig:burstrate}
  as a function of the time to failure $t_c-t$ for the phase of finite lifetime $A<A_c^*$. 
  The approach to failure is characterized by an Omori type acceleration. The bold black line
  represents the Omori law Eq.\ (\ref{eq:omori}).
  \label{fig:omori}
}
\end{figure}
It can be observed in Fig.\ \ref{fig:omori} that as the event rate increases 
when failure is approached, it gets saturated close to $t_c$. 
The functional form of the curves 
can be well approximated by the Omori law of earthquakes
\beq{
\left<n_{\Delta}\right>(t) = \frac{K}{\left(1+\frac{t_c-t}{t_0} \right)^p},
\label{eq:omori}
}
where $K$ denotes the saturation rate, $t_0$ is the characteristic time distance 
from $t_c$ where saturation sets in, and $p$ is the exponent, which controls 
the power law decay beyond $t_0$. Simulations
performed varying the parameters $\sigma_0$, $A$, and $\gamma$ in broad ranges revealed 
that the exponent $p$ exhibits a high degree of universality, i.e. best fit of the data is provided 
by the Omori exponent $p\approx 1$ in all cases.
It is interesting to note that for earthquakes the Omori law describes the relaxation of the system
following major shocks, while in our case it provides a good quality quantitative description 
of the acceleration of the system towards macroscopic failure. The value of our Omori exponent
falls close to the typical value of earthquakes \cite{ojala_2004}.

\section{Forecasting global failure}
It is a crucial question whether the imminent global failure of the bundle can be forecasted.
In order to foresee failure, signatures have to be identified early enough, which allow for
a quantitative prediction of the time of final collapse $t_c$. 
Breaking events induced by slow damage accumulation can hardly be monitored in reality 
\cite{lockner_1993,sabine_PhysRevE.88.032802},
however, avalanches of fiber breakings are sudden outbreaks which generate acoustic waves
\cite{meinders_scaling_2008,rosti_statistics_2010,salje_main_minecollapse_2017}.
Based on acoustic emission measurements the fracture process can be decomposed as a
time series of discrete fracturing events, which correspond to the breaking avalanches 
of our model \cite{PhysRevLett.110.088702,vives_0953-8984-25-29-292202}.

\begin{figure}
\begin{center}
\epsfig{bbllx=0,bblly=0,bburx=350,bbury=310,file=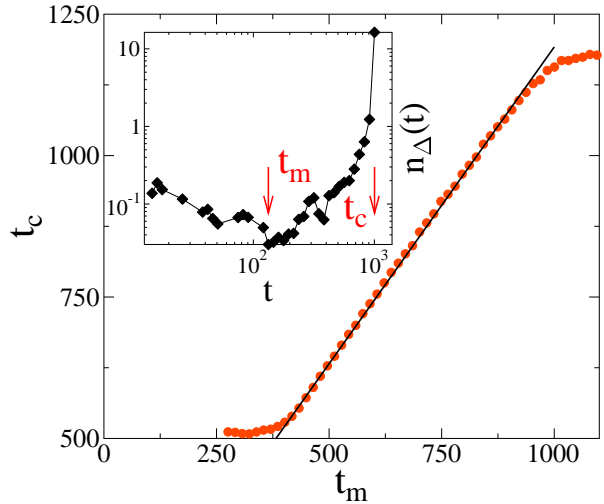, width=8.3cm}
\end{center}
  \caption{{\it (Color online)} Average failure time as a function of $t_{m}$. 
  The value of $t_{m}$ was binned and for each bin the average of $t_c$ was calculated, 
  and finally plotted   as a function of the average $t_{m}$ within one bin.
  Inset: Rate of bursts $n_{\Delta}$ in a single sample 
  as a function of time $t$ obtained at the unloading rate $A/A_c^*=0.997$. 
  The event rate takes a minimum value at time $t_m$.
  Global failure occurs at time $t_c$ preceeded by an acceleration with a 
  rapidly increasing event rate.
  \label{fig:tctmin}
}
\end{figure}

We have seen above that in the phase of finite lifetime global failure is preceded 
by a short acceleration period, which sets on after the time  $t_{m}$ of minimum event 
rate has been passed. 
The inset of Fig.\ \ref{fig:tctmin} demonstrates that the value of $t_{m}$ can be measured 
with a good confidence even in a single system without averaging. 
It follows that in case a correlation exists between the lifetime $t_c$ 
and the time of minimum event rate $t_{m}$, the value of $t_{m}$ could be used 
to predict the failure time $t_c$. 

We performed computer simulations at a fixed system size of $N=10^6$ fibers and determined
the two characteristic times $t_{m}$ and $t_c$ for a large number of individual systems
at several unloading rates $A$. 
To quantify the correlation of the two time scales we binned
$t_{m}$ irrespective of the corresponding unloading rate $A$, 
and calculated the average failure time for each bin.
In Fig.\ \ref{fig:tctmin} the average failure time is plotted as a function
of the average minimum time. The correlation of the two quantities can be very well described
by a linear functional form
\beq{
t_c = B t_{m}, 
\label{eq:tctmin}
}
where the multiplication factor was obtained by fitting $B=1.12\pm 0.005$.
The result implies that when the event rate has passed its minimum 
then the failure time can be estimated according to the relation Eq.\ 
(\ref{eq:tctmin}). The value of $B$ falling close to 1 shows that the 
time evolution of the system is strongly asymmetric, the acceleration 
period is only about 10 percent of the duration of the slowdown. This 
short acceleration period may still be sufficient to mitigate consequences 
of the final collapse of the system. 

\begin{figure}
\begin{center}
\epsfig{bbllx=30,bblly=30,bburx=375,bbury=335,file=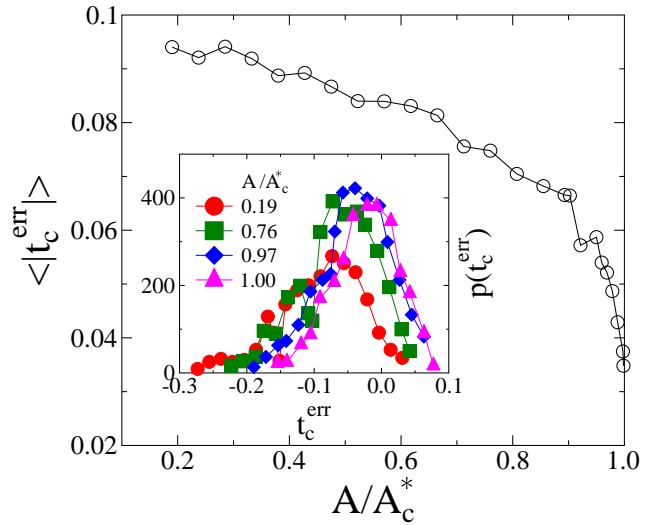, width=8.3cm}
\end{center}
  \caption{{\it (Color online)} The average of the absolute value of the relative error 
  $\left<|t_c^{err}|\right>$ of lifetime estimates as a function of the unloading rate $A$. 
  Approaching the critical point the quality of forecasting rapidly improves.
  Inset: Probability distribution $p(t_c^{err})$ of the relative error $t_c^{err}$ of lifetime estimates
  for several values of $A$.
  \label{fig:error}
}
\end{figure}

To quantify the reliability of the 
forecasting method, we determined the probability distribution $p(t_c^{err})$ 
of the relative error $t_c^{err}$ of the lifetime estimates defined as 
\beq{
t_c^{err} = \frac{Bt_m - t_c}{t_c}.
}
Here $t_c$ is the measured lifetime of a single sample and $B=1.12$ was substituted to
make an estimate based on the corresponding $t_m$. 
The inset of Fig.\ \ref{fig:error} presents the distribution $p(t_c^{err})$ 
for several values of the control parameter $A$ where each curve was obtained using 1000 samples.
It can be observed that $p(t_c^{err})$ is asymmetric and for $A$ values far from
the critical point $A_c^*$ even the most probable error is negative. The result implies
that in this parameter range the method underestimates the lifetime of samples with a relatively 
broad scatter.
However, approaching the critical unloading rate $A_c^*$ the distributions get narrower
and shift towards positive values, which indicate the improvement of forecasting.
We also determined the average of the absolute value of the relative error $\left<\left|t_c^{err}\right|\right>$ as a function of $A$. Figure \ref{fig:error} shows that 
far from the critical unloading rate $A_c^{*}$ the relative error is above 10\%.
Since the acceleration period has a comparable duration relative to the lifetime
of the sample, the reliability of forecasting is low in this parameter range.
However, in the vicinity of the critical point the quality of estimates
rapidly improves, since the relative error drops down to a few procent.

Similar forecasting methods have been suggested for creep 
rupture processes under a constant external load 
\cite{monkman_empirical_1956,sornette_predictability_2002,
kovacs_critical_2008,alava_lifetime_pre2016}, but these alternatives 
mainly focus on specific features of the evolution of the deformation rate 
$\dot{\varepsilon}$
as a function of time $t$. The Monkman-Grant (MG) relation 
of materials expresses the sample creep life $t_c$ as a power law of the 
minimum strain rate $\dot{\varepsilon}_{min}$ \cite{monkman_empirical_1956}. 
Although the MG relation
was determined on an empirical ground, for some cases it has been derived 
analytically from models of damage enhanced creep such as the visco-elastic 
fiber bundle model \cite{kovacs_critical_2008}. A linear relation between 
the lifetime $t_c$ of creeping samples and the time of minimum strain 
rate $t_m$ was pointed out,  similar to Eq.\ (\ref{eq:tctmin}) for fiber composites. 
Careful experiments revealed a universality of the multiplication factor 
$t_c\approx3/2t_m$ significantly larger than our one 
\cite{nechad_andrade_2005,nechad_creep_2005}. Recently, a similar 
relation was suggested to describe the lifetime of loaded paper sheets
where the creep process proved to be more anisotropic with $t_c\approx1.2t_m$ 
\cite{alava_lifetime_pre2016}.

\section{Discussion}
Based on a model of damage enhanced creep rupture we investigated 
the time evolution of the rupture process emerging under unloading
from an initial load. 
When fiber breaking is only induced by aging, the time evolution of the 
system can be explored by analytical calculations.
We showed analytically that the system has two phases: for slow unloading global failure occurs
in a finite time, while rapid unloading gives rise to partial failure 
and an infinite lifetime of the bundle. The transition between the two phases
occurs at a critical unloading rate analogous to continuous phase transitions. 
The critical unloading rate depends on the parameters of the damage law of the 
model such that at higher initial loads a faster unloading is required to prevent
failure. 
The phase structure of the system can be characterized by the fraction of 
intact fibers when unloading is completed playing the role of the order 
parameter of the system. This quantity is identically zero
in the phase of finite lifetime, while it takes a finite non-zero value 
when partial failure occurs. Approaching the transition from the phase of 
partial failure, the order parameter goes to zero according to a power law of 
the distance from the critical point.
Analytical calculations revealed that the order 
parameter exponent of the transition is not universal in the sense that 
it depends on the exponent of the damage law of the model.

When both breaking mechanisms are active, a highly complex fracture 
process emerges, where slowly proceeding damage sequences trigger 
bursts of immediate breakings. 
Computer simulations revealed that avalanche triggering does not change the 
overall character of the system, however, the rupture process gets faster
and the critical unloading rate shifts to a higher value.
Based on finite size scaling of the simulation results,
we determined the asymptotic critical point
of the system in the limit of infinite bundle size.
The beginning of the unloading process
is accompanied by a relaxation of the bursting activity quantified by a power
law decrease of the event rate. 
In the phase of finite lifetime, macroscopic failure is preceded
by a short acceleration period described by the (inverse) Omori law.
The Omori exponent of our fracture process has the same value as 
the one of earthquakes. The critical exponents characterizing the bursting
activity of the rupture process proved to be universal, they are only affected
by the range of load sharing between fibers.

We pointed out a simple linear
relation between the time where the burst rate reaches its minimum and the 
lifetime of the sample, which can be exploited to forecast the imminent 
catastrophic failure of the system. Simulations showed that the forecasting 
method has a good reliability in the vicinity of the critical unloading rate.
In laboratory or field measurements 
using the evolution of the time series of acoustic outbreaks 
as information source suggested by our results, may be advantageous compared to
monitoring the strain rate used by other methods of forecasting.

It is an important characteristics of our model that the external load is always below the 
fracture strength of the system, and hence, failure is driven by the slow aging 
of loaded fibers even when both breaking mechanisms are active. 
It follows that at any constant external load the bundle would 
have a finite lifetime. Unloading from an initial load gives rise to a competition 
of two mechanisms: 
the decreasing external load results in slowdown of the time evolution, 
while fiber breaking reduces the load 
bearing capacity which accelerates the fracture process. The competition is controlled
by the unloading rate in such a way that sufficiently fast unloading can prevent 
global failure. Such loading situations may arise in geological systems,
when e.g.\ the direction of the motion of tectonic plates gets reversed giving rise 
to slow unloading of plate boundaries.

Although, our study only considered the case of a linearly decreasing external load, the 
application of the model to more complex time dependences, including also cyclic loading, 
is straightforward. As the simplest extension of our calculations, it is worth analysing 
how the system evolves when the external load linearly increases. In this case, the bundle
always fails globally, however, as the loading rate increases the slow damage mechanism
plays a diminishing role, and eventually the fracture process is entirely controlled by the immediate 
breaking of fibers. Cyclic loading is composed of loading and unloading periods, where an 
interesting competion of damage and immediate breaking can be expected. Work on this 
is in progress.

\begin{acknowledgments}
The project is co-financed by the European Union and the European Social Fund.
The work is supported by the EFOP-3.6.1-16-2016-00022 project. 
This research was supported by the National Research, Development and
Innovation Fund of Hungary, financed under the K-16 funding scheme Project no.\ K 119967.
R.K.\ was supported by the \'UNKP-17-2 New National 
Excellence Program of the Ministry of Human Capacities”.
The research was financed by the Higher Education Institutional
Excellence Programme of the Ministry of Human Capacities in Hungary, 
within the framework of the Energetics thematic
programme of the University of Debrecen.
\end{acknowledgments}

\bibliography{/home/feri/papers/statphys_fracture}

\end{document}